\documentclass[twocolumn,showpacs,amsmath,amssymb,superscriptaddress,floatfix]{revtex4}
\usepackage{graphicx}%
\usepackage{slashed}

\begin{document}
\title{Associated strangeness production in the $\boldsymbol{pp \to pK^+K^-p}$
and $\boldsymbol{pp \to pK^+\pi^0\Sigma^0}$ reactions}
\author{Ju-Jun \surname{Xie}} \email{xiejujun@ific.uv.es}
\affiliation{Instituto de Fisica Corpuscular (IFIC), Centro Mixto
CSIC-Universidad de Valencia, Institutos de Investigacion de Paterna,
Aptd. 22085, E-46071 Valencia, Spain}%
\affiliation{Department of Physics, Zhengzhou University, Zhengzhou,
Henan 450001, China}
\author{Colin Wilkin}\email{cw@hep.ucl.ac.uk}
\affiliation{Physics and Astronomy Department, UCL, London WC1E 6BT,
United Kingdom}
\date{\today}%

\begin{abstract}
The total and differential cross sections for associated strangeness
production in the $pp \to pK^+K^-p$ and $pp \to pK^+\pi^0\Sigma^0$
reactions have been studied in a unified approach using an effective
Lagrangian model. It is assumed that both the $K^-p$ and
$\pi^0\Sigma^0$ final states originate from the decay of the
$\Lambda(1405)$ resonance which was formed in the production chain
$pp\to p(N^*(1535)\to K^+\Lambda(1405))$. The available experimental
data are well reproduced, especially the ratio of the two total cross
sections, which is much less sensitive to the particular model of the
entrance channel. The significant coupling of the $N^*(1535)$
resonance to $\Lambda(1405) K$ is further evidence for large $s
\bar{s}$ components in the quark wave function of the $N^*(1535)$
resonance.

\end{abstract}
\pacs{13.75.-n.; 14.20.Gk.; 13.30.Eg.} \maketitle

\section{Introduction}

The $N^*(1535)$ isobar has proved to be a controversial resonance for
many years. In the simple three-quark constituent model, the odd
parity $N^*(1535)(J^p = 1/2^-)$ should be the lowest spatially
excited nucleon state, with one quark in a $p-$wave. However, the
even parity $N^*(1440)$ has in fact a much lower mass, despite
requiring two units of excitation energy. This is the long-standing
mass inversion problem of the nucleon spectrum.

The $N^*(1535)$ resonance couples strongly to the $\eta N$
channel~\cite{pdg2008} but a large $N^*(1535)K \Lambda$ coupling has
also been deduced~\cite{garciaplb582,liuprl96,gengprc79} through the
analysis of BES data on $J/\psi \to p\bar{p} \eta, \bar{p}\Lambda
K^+$ decays~\cite{BES} and COSY data on the $pp \to p\Lambda K^+$
reaction near threshold~\cite{COSY11}. Analyses~\cite{Mosel,Saghai}
of recent SAPHIR~\cite{ELSA} and CLAS~\cite{CLAS} $\gamma p\to
K^+\Lambda$ data also indicate a large coupling of the $N^*(1535)$ to
$K\Lambda$.

In a chiral unitary coupled channel model it is found that the
$N^*(1535)$ resonance is dynamically generated, with its mass, width,
and branching ratios in fair agreement with
experiment~\cite{garciaplb582,osetprc65,kaisernpa612,nievesprd64,doeringepja43}.
This approach shows that the couplings of the $N^*(1535)$ resonance
to the $K \Sigma$, $\eta N$ and $K \Lambda$ channels could be large
compared to that for $\pi N$. Data on the $\gamma p \to p
\eta'$~\cite{dugger} and $pp \to pp\eta'$ reactions~\cite{caoxuprc78}
suggest also a coupling of the isobar to $\eta' N$. In addition,
there is some evidence for a $N^*(1535)N\phi$ coupling from the
$\pi^- p \to n\phi$ and $pp \to pp \phi$~\cite{doringprc78,xieprc77}
as well as the $pn \to d\phi$~\cite{caoxuprc80} reactions.

The mass inversion problem could be understood if there were a
significant $s\bar{s}$ components in the $N^*(1535)$ wave
function~\cite{zoureview,zhangan} and this would also provide a
natural explanation of its large couplings to the strangeness
$K\Lambda$, $K\Sigma$ $N\eta'$ and $N\phi$ channels. It would
furthermore lead to an improvement in the description of the helicity
amplitudes in $N^*(1535)$ photoproduction~\cite{anepja39}. We wish to
argue in this paper that a hidden strangeness component in the
$N^*(1535)$ might play a much wider role in associated strangeness
production in medium energy nuclear reactions.

The $\Lambda(1405)(J^p = 1/2^-)$ can be considered as the strangeness
$S=-1$ counterpart of the $N^*(1535)$ and its structure is possibly
even more controversial. In quark model calculations, it is described
as a $p-$wave $q^3$ baryon~\cite{isgurprd18} but it can also be
explained as a $\bar{K}N$ molecule~\cite{dalitz} or $q^4\bar{q}$
pentaquark state~\cite{inouenpa790}. On the other hand, within
unitary chiral theory~\cite{kaisernpa612,garciaplb582,chiral}, two
overlapping $I=0$ states are dynamically generated and in this
approach the shape of any observed $\Lambda(1405)$ spectrum might
depend upon the production process. In a recent study of the $pp\to
pK^+\Lambda(1405)$ reaction~\cite{zychorplb660} the resonance was
clearly identified through its $\pi^0\Sigma^0$ decay and no obvious
mass shift was found. However, this result is inconclusive in the
sense that the data could also be well described in the two-resonance
scenario~\cite{gengepja34}. For simplicity we shall here work within
the single $\Lambda(1405)$ framework with parameters as reported in
the PDG review~\cite{pdg2008}.

In parallel with the $\Lambda(1405)$ measurement, Maeda \emph{et
al.}\ also extracted differential and total cross sections for kaon
pair production in the $pp \to pp K^+ K^-$
reaction~\cite{maedaprc77}. These results show clear evidence for the
excitation and decay of the $\phi$ meson sitting on a smooth $K^+K^-$
background, whose shape resembles phase space. It has been
suggested~\cite{WIL09} that the $\Lambda(1405)$ could be important
for the non-$\phi$ kaon pair production through the $pp \to pK^+
(\Lambda(1405)\to K^-p)$ reaction. This would, of course, only be
relevant for the isospin $I=0$ $K^-p$ contribution but this is likely
to dominate the low mass region because of the presence of the
$\Lambda(1405)$. It is therefore the purpose of the present paper to
analyze simultaneously the available data on $pp\to
K^+p\Sigma^0\pi^0$ and $pp\to K^+pK^-p$ production at a beam energy
of 2.83~GeV~\cite{zychorplb660,maedaprc77} within a unified
phenomenological model, where the $N^*(1535)$ isobar acts as a
doorway state for both production processes.

The foundation of the model is the assumption that there are large
$s\bar{s}$ components in the quark wave function of the $N^*(1535)$
isobar and that these induce a significant $N^*(1535):\Lambda(1405)K$
coupling. This in turn allows the possibility that the production of
the $\Lambda(1405)$ in proton-proton and $\pi^- p$ collisions could
be dominated by the excitation and decay of the $N^*(1535)$ resonance
below the $\Lambda(1405) K$ threshold. Within this picture, we
calculate the $pp \to pK^+ \Lambda(1405) \to pK^+(K^-p/\pi^0
\Sigma^0)$ and $\pi^- p \to \Lambda(1405) K^0$ reactions using an
effective Lagrangian approach. We show that the pion-induced data are
indeed compatible with the large $N^*(1535):\Lambda(1405)K$ coupling
resulting from the $s \bar{s}$ components in the $N^*(1535)$. The
resulting theoretical estimates of the $pp \to pK^+ K^-p$ and $pp \to
pK^+ \pi^0 \Sigma^0$ differential and total cross sections describe
well the available COSY experimental
data~\cite{zychorplb660,maedaprc77}. In particular, the ratio of
these two cross sections, where many of the theoretical uncertainties
cancel, is reproduced within the total theoretical and experimental
uncertainties.

Section~\ref{formalism} presents the formalism and ingredients
required for the calculation, with the numerical results and
discussions being given in Sec.~\ref{results}. A short summary and a
presentation of our conclusions then follows in
Sec.~\ref{conclusions}

%
%

\section{Formalism and ingredients}
\label{formalism}

We study the $pp \to pK^+ \Lambda(1405) \to pK^+(K^-p/\pi^0
\Sigma^0)$ and $\pi^- p \to \Lambda(1405) K^0$ reactions in an
effective Lagrangian approach on the assumption that the production
of the $K\Lambda(1405)$ pair is dominantly through the excitation and
decay of the sub-threshold $N^*(1535)$ resonance. It is generally
assumed that the production of $\eta$ mesons in nucleon-nucleon
collisions near threshold passes mainly through the $N^*(1535)$,
which has a very strong coupling to $N\eta$. However, there is far
from unanimity in the modelling of these processes within a
meson-exchange picture, with different groups considering $\pi$,
$\rho$, $\eta$, and $\omega$ exchanges to be
important~\cite{xieprc77,models}. Fortunately, the estimation of the
$pp \to pK^+ \Lambda(1405)$ cross section in our model is only
sensitive to the production rate of the $N^*(1535)$ and single pion
exchange is sufficient for this purpose. By neglecting $\eta$ and
$\rho$ exchanges, we can present a unified picture of pion- and
proton-induced production processes, though our results are more
general than this would suggest.

\begin{figure}[htdp]
\includegraphics[scale=0.5]{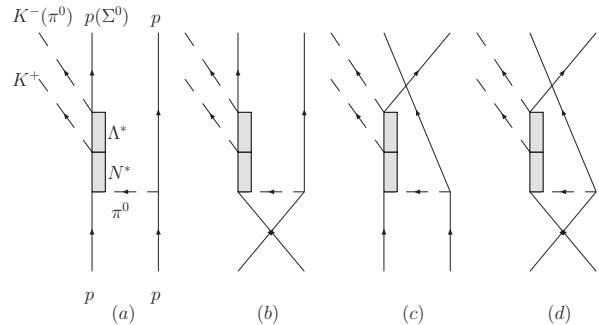}
\vspace{-0.2cm} \caption{Feynman diagrams for the $pp \to pK^+
\Lambda(1405) \to pK^+ (K^-p/\pi^0\Sigma^0)$ reaction.}
\label{diagram}
\end{figure}

\begin{figure}[htdp]
\includegraphics[scale=0.80]{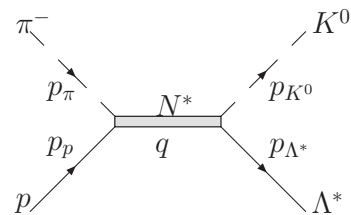}
\vspace{-0.2cm} \caption{Feynman diagram for $\pi^- p \to
\Lambda(1405) K^0$ reaction. Here $p_{\pi}$, $p_p$, $p_{K^0}$,
$p_{\Lambda^*}$ and $q$ are the four-momenta of $\pi^-$, proton,
$K^0$, $\Lambda(1405)$, and intermediate $N^*(1535)$ resonance,
respectively.} \label{pipdiagram}
\end{figure}

The basic Feynman diagrams for the $t$-channel exchanges in $pp \to
pK^+ \Lambda(1405) \to pK^+ (K^-p/\pi^0 \Sigma^0)$ reaction and the
$s-$channel diagram for $\pi^- p \to \Lambda(1405) K^0$ are depicted
in Figs.~\ref{diagram} and \ref{pipdiagram}, respectively. For the
$pp \to pK^+\pi^0\Sigma^0$ reaction, only diagrams in
Figs.~\ref{diagram}(a) and \ref{diagram}(b) need to be considered,
while for the $pp \to pK^+ K^- p$ reaction, the exchange terms
\ref{diagram}(c) and \ref{diagram}(d) have also to be included.

We employ the commonly used interaction Lagrangian for the $\pi NN$
vertex,
\begin{equation}
\mathcal{L}_{\pi N N}  = -i g_{\pi N N} \bar{N} \gamma_5 \vec\tau
\cdot \vec\pi N\,, \label{pinn}
\end{equation}
with an off-shell form factor taken from the Bonn potential
model~\cite{mach}
\begin{equation}
F^{NN}_{\pi}(k^2_{\pi}) =
\frac{\Lambda^2_{\pi}-m_{\pi}^2}{\Lambda^2_{\pi}- k_{\pi}^2},
\label{formf}
\end{equation}
where $k_{\pi}$, $m_{\pi}$ and $\Lambda_{\pi}$ are the
four-momentum, mass and cut-off parameter for the exchanged pion.
The coupling constant and the cutoff parameter are taken to be
$g_{\pi NN}^2/4\pi = 14.4$ and $\Lambda_{\pi} = 1.3$
GeV/$c^2$~\cite{mach,tsushima}.

To evaluate the invariant amplitudes corresponding to the diagrams of
Figs.~\ref{diagram} and \ref{pipdiagram}, we also need to know the
interaction Lagrangians involving the $N^*(1535)$ and $\Lambda(1405)$
resonances. In Ref.~\cite{zouprc03}, a Lorentz-covariant orbital-spin
($L$-$S$) scheme for $N^* N M$ couplings was studied in detail.
Within this approach, the $N^*(1535)N\pi$, $N^*(1535)\Lambda(1405)K$,
$\Lambda(1405)\bar{K}N$ and $\Lambda(1405)\pi \Sigma$ effective
couplings become,
\begin{eqnarray}
\nonumber \mathcal{L}_{N^* N \pi }  &=&  -ig_{N^* N\pi}\bar{N^*}
\vec\tau \cdot
\vec\pi N + \text{h.c.},\\
\nonumber \mathcal{L}_{N^*\Lambda^* K} &=& \frac{g_{N^* \Lambda^*
K}}{m_{K}} \bar{N^*}\gamma_5\gamma_{\mu}
\partial^{\mu}K \Lambda^* + \text{h.c.}, \\
\nonumber \mathcal{L}_{\Lambda^* \bar{K} N} &=& -i g_{\Lambda^*
\bar{K} N}
\bar{N} \bar{K} \Lambda^* +\text{h.c.}, \\
\mathcal{L}_{\Lambda^* \pi \Sigma} &=& -ig_{\Lambda^* \pi \Sigma}
\bar{\Lambda^*} \vec\pi  \cdot \vec\Sigma +
\text{h.c.}.\label{pinnstar}
\end{eqnarray}

To minimize the number of free parameters, a similar dipole form
factor to that of Eq.~\eqref{formf} will be used for the $N^*(1535) N
\pi$ vertex, with the same value of the cut-off parameter.

The $N^*(1535)N\pi$, and $\Lambda(1405)\pi\Sigma$ coupling constants
are determined from the partial decay widths of these two
resonances~\cite{pdg2008}. With the effective interaction Lagrangians
of Eqs.~\eqref{pinnstar}, the coupling constants are related to the
partial decay widths by
\begin{eqnarray}
\nonumber \Gamma_{N^*(1535) \to N \pi} &=& \frac{3 g^2_{N^*N
\pi}(m_N+E_N)p^{\,\text{cm}}_N}{4\pi M_{N^*}}, \label{1535pi}
\end{eqnarray}
where
\begin{eqnarray}
\nonumber p^{\,\text{cm}}_N &=&
\frac{\lambda^{1/2}(M^2_{N^*},m^2_N,m^2_{\pi})}{2M_{N^*}},
\end{eqnarray}
with
\begin{equation}
E_N = \sqrt{(p^{\,\text{cm}}_N)^2+m^2_N},
\end{equation}
and correspondingly for the $\Lambda(1405)\to \pi\Sigma$ decay in
terms of the $g_{\Lambda^*\pi\Sigma}$ coupling constant. Here
$\lambda$ is the triangle function,
\begin{eqnarray}
\lambda(x,y,z)=x^2+y^2+z^2-2xy-2xz-2yz.
\end{eqnarray}

Although the mass differences do not allow one to obtain directly
similar results for the $N^*(1535) \Lambda(1405) K$ vertex, the
requisite information can be extracted from $\pi^- p \to
\Lambda(1405) K^0$ data, provided that this reaction is dominated by
the $s$-channel $N^*(1535)$ pole of Fig~\ref{pipdiagram}. The
corresponding invariant amplitude $\mathcal{A}$ becomes:
\begin{eqnarray}
\mathcal{A} &=& \frac{\sqrt{2} g_{N^* N \pi} g_{N^* \Lambda^*
K}}{m_K} F_{N^*}(q^2) \bar{u}(p_{\Sigma},s_{\Sigma})  \times
\nonumber \\
&& \gamma_5\slashed{p}_{K^0} G_{N^*}(q) u(p_p,s_p),
\end{eqnarray}
where $s_{\Sigma}$ and $s_p$ are the baryon spin projections.

The form factor for the $N^*(1535)$ resonance, $F_{N^*}(q^2)$, is
taken in the form advocated in Refs.~\cite{Mosel, feuster}:
\begin{equation}
F_{N^*}(q^2)=\frac{\Lambda_{N^*}^{\:4}}{\Lambda_{N^*}^{\:4} +
(q^2-M^2_{N^*})^2},
\end{equation}
with $\Lambda_{N^*} = 2.5$~GeV/$c^2$.

The $N^*(1535)$ propagator is written in a Breit-Wigner
form~\cite{liang}:
\begin{equation}
G_{N^*}(q)=\frac{i(\slashed{q}
+M_{N^*})}{q^2-M^2_{N^*}+iM_{N^*}\Gamma_{N^*}(q^2)}\,,
\end{equation}
where $\Gamma_{N^*}(q^2)$ is the energy-dependent total width.
Keeping only the dominant $\pi N$ and $\eta N$ decay
channels~\cite{pdg2008}, this can be decomposed as
\begin{equation}
\Gamma_{N^*} (q^2) = a_{\pi N}\, \rho_{\pi N}(q^2) + b_{\eta N} \,
\rho_{\eta N}(q^2),
\end{equation}
where $a_{\pi N} = 0.12$~GeV/$c^2$, $b_{\eta N} = 0.32$~GeV/$c^2$,
and the two-body phase space factors, $\rho_{\pi(\eta)N}(q^2)$, are
\begin{equation}
\rho(q^2)= \left.2
p^{\,\text{cm}}(q^2)\,\Theta(q^2-q_{\text{thr}}^2)\right/\!\!\sqrt{q^2}\,,
\label{psf}
\end{equation}
and $q_{\text{thr}}$ is the threshold value for the decay
channel.

A similar representation is adopted for the $\Lambda(1405)$
propagator and form factor, with the same value of the cut-off
parameter $\Lambda_{\Lambda^*} = 2.5$~GeV/$c^2$. Because the
$\Lambda(1405)$ resonance lies slightly below the $\bar{K}N$
threshold, the only nominally allowed decay channel is $\pi \Sigma$.
Nevertheless, ever since the pioneering work of Dalitz and
Tuan~\cite{dalitz} it has been known that there is also a strong
coupling to $\bar{K}N$. The ensemble of low energy data on $K^-p$ and
related channels has been described in terms of a separable potential
model~\cite{gal}. In contrast to the unitary chiral
approach~\cite{gengepja34}, the separable model produces only a
single $\Lambda(1405)$ pole and from this we can investigate its
effects above the $\bar{K}N$ threshold. These can be parametrized in
terms of an energy dependent partial width
\begin{equation}
\Gamma_{\Lambda(1405)} (q^2) = a_{\pi \Sigma}\,\rho_{\pi\Sigma}(q^2)
+ b_{\bar{K} N}\,\rho_{\bar{K} N}(q^2)\,, \label{wid1405}
\end{equation}
where $a_{\pi \Sigma} = 0.22$~GeV/$c^2$, $b_{\bar{K}N} =
0.49$~GeV/$c^2$ and the two-body phase space factors are given in
Eq.~(\ref{psf}). By using
\begin{eqnarray}
g^2_{\Lambda^*\bar{K}N} = \frac{0.49 \times 3 \times
(m_{\Sigma}+E_{\Sigma}(q^2))}{0.22 \times 2 \times (m_N+E_N(q^2))}
\times g^2_{\Lambda^* \pi \Sigma}\,, \label{gkn}
\end{eqnarray}
the width equation \eqref{wid1405} leads to a $\Lambda(1405)\bar{K}N$
coupling constant $g^2_{\Lambda^*\bar{K}N}/4\pi =0.27$ at the
$\bar{K}N$ threshold.

\begin{figure}[htdp]
\includegraphics[scale=0.4]{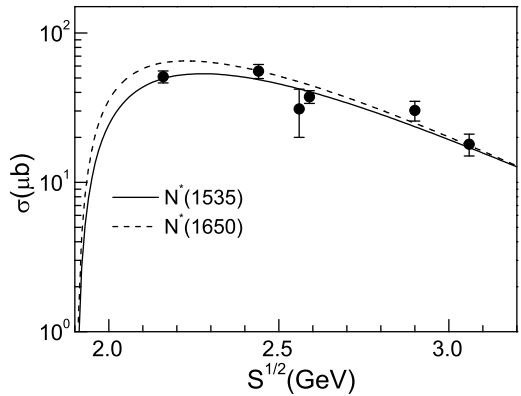}%
\vspace{-0.50cm} \caption{ Total cross section for the $\pi^- p \to
\Lambda(1405) K^0$ reaction as a function of the c.m.\ energy
$\sqrt{s}$. The solid curve represents the fit of the $s$-channel
$N^*(1535)$ pole of Fig.~\ref{pipdiagram} to the available
experimental data~\cite{pipdata}. The dashed curve is the
corresponding fit if the $N^*(1650)$ resonance were used instead.}
\label{piptcs}%
\end{figure}

We now evaluate the $\pi^- p \to \Lambda(1405) K^0$ total cross
section as a function of the center-of-mass energy. The value of the
$N^*(1535) \Lambda(1405) K$ coupling constant
$g^2_{N^*\Lambda^*K}/4\pi = 0.28$ leads to the predictions that are
compared with experimental data~\cite{pipdata} in Fig.~\ref{piptcs}.
Although the agreement is reasonable, it must be stressed that the
predictions are not very sensitive to the mass of the $N^*$, provided
it lies well below the $K\Lambda(1405)$ threshold. As can be judged
from the figure, a very similar shape would be obtained if one used
for example the second $S_{11}$ resonance $N^*(1650)$. However, it
has been shown~\cite{xieprc77} that a large $s\bar{s}$ component in
the $N^*(1650)$ resonance is not consistent with its smaller coupling
to $N\eta$ than $N\pi$. It should also be noticed that any possible
contributions from $t$- and $u$-channel exchanges have also been
neglected. The value of this coupling constant is given along with
others in Table.~\ref{tab1}.

\begin{table}[htdp]
\caption{\label{table} Values of the coupling constants required for
the estimation of the $pp \to pK^+ K^-p$ and $pp \to pK^+ \pi^0
\Sigma^0$ cross sections. These have been estimated from the
branching ratios quoted~\cite{pdg2008}, though it should be noted
that these are for all final charged states. As described in the
text, the $\Lambda^* \bar{K} N$ coupling was obtained from the
energy dependence of the $\Lambda(1405)$ width given by
Eq.~(\ref{wid1405}), and the $N^* \Lambda^* K$ coupling was derived
from measurements of the $\pi^- p \to \Lambda(1405) K^0$ total cross
section.}
\begin{center}
\begin{tabular}{ccc}
\hline Vertex  & Branching ratio & $g^2/4\pi$\\
\hline $N^* N \pi$ & $0.45$ & $0.038$ \\
       $\Lambda^* \pi \Sigma$ & $1.00$ & $0.064$ \\
       $\Lambda^* \bar{K} N$ & --- & $0.27$ \\
       $N^* \Lambda^* K$ & ---  & $0.28$ \\
\hline
\end{tabular}
\end{center} \label{tab1}
\end{table}

The full invariant amplitude for the $pp \to pK^+ K^- p$ reaction is
composed of four parts, corresponding to the diagrams shown in
Fig.~\ref{diagram};
\begin{eqnarray}
\mathcal{M} = \sum_{i = a, b, c, d} \eta_i\,\mathcal{M}_i.
\label{amp}
\end{eqnarray}
To take account of the antisymmetry of the protons in the initial and
final states, factors $\eta_a = \eta_d =1$ and $\eta_b = \eta_c = -1$
are introduced. It is important to note that only $\mathcal{M}_a$ and
$\mathcal{M}_b$ should be considered for the $pp \to p K^+ \pi^0
\Sigma^0$ reaction.

Each amplitude can be derived straightforwardly with the effective
couplings given. We give as an example the form of the
$\mathcal{M}_a$ amplitude:
\begin{eqnarray}
\mathcal{M}_a  &=& \frac{g_{\pi NN} g_{N^* N \pi} g_{N^* \Lambda^* K}
g_{\Lambda^* \bar{K} N}}{m_K} F^{N N}_{\pi}(k^2_{\pi}) \times \nonumber \\
&&F^{N^* N}_{\pi}(k^2_{\pi}) F_{N^*}(q_1^2) F_{\Lambda^*}(q_2^2)
G_{\pi}(k_{\pi}) \bar{u} (p_{\,4},s_4) \times  \nonumber \\
&& G_{\Lambda(1405)}(q_2) \gamma_5 \slashed{p}_{\,5}
G_{N^*(1535)}(q_1) u(p_{\,1},s_1) \times  \nonumber \\ &&
\bar{u}(p_{\,3},s_3) \gamma_5 u(p_{\,2},s_2),
\end{eqnarray}
where $s_i~(i=1,2,3,4)$ and $p_i~(i=1,2,3,4)$ represent the spin
projections and four-momenta of the two initial and two final
protons, respectively. The $q_1$ and $q_2$ are the four-momenta of
intermediate $N^*(1535)$ and $\Lambda(1405)$ resonances, while
$p_{\,5}$ is the four-momentum of the final $K^+$ meson. The pion
propagator is
\begin{equation}
G_{\pi}(k_{\pi})=\frac{i}{k_{\pi}^2-m^2_{\pi}}\,\cdot
\end{equation}

The final-state-interaction(FSI) between the two emerging protons in
the $^{1\!}S_0$ wave in the $pp\to ppK^+K^-$ case is taken into
account using the Jost function formalism~\cite{gill}, with
\begin{eqnarray}
J(q)^{-1}=\frac{k+ i \beta}{k- i \alpha}, \label{fsi}
\end{eqnarray}
where $k$ is the internal momentum of $pp$ subsystem. The parameters
$\alpha = -20.5$~MeV$/c$ and $\beta =
166.7$~MeV$/c$~\cite{sibiepja06} give a slightly stronger $pp$ FSI in
the near-threshold region than that used in the experimental
paper~\cite{maedaprc77}.

The normalization is chosen such that the differential cross section
is
\begin{eqnarray}
&& d\sigma (pp \to pK^+ pK^-) = \frac{m^2_p}{F}
\frac{1}{4}\sum_{s_i,s_f} |\mathcal{M}|^2  \frac{m_p d^{3}
p_{3}}{E_{3}}\times \nonumber \\
&& \hspace{-8mm}\frac{m_p d^{3} p_4}{E_4} \frac{d^{3} p_5}{2 E_5}
\frac{d^{3} p_6}{2 E_6} \frac{1}{2}\,
\delta^4(p_1+p_2-p_3-p_4-p_5-p_6), \label{eqcs}
\end{eqnarray}
with the flux factor
\begin{eqnarray}
F=(2 \pi)^8\sqrt{(p_1\cdot p_2)^2-m^4_p}\,. \label{eqff}
\end{eqnarray}

The factor $\frac{1}{2}$ before the $\delta$-function in
Eq.~\eqref{eqcs} results from having two final identical protons and
must be omitted for the $pK^+ \pi^0 \Sigma^0$ final state.

%
%

\section{Numerical results and discussion}
\label{results}

\begin{figure}[htdp]
\includegraphics[scale=0.4]{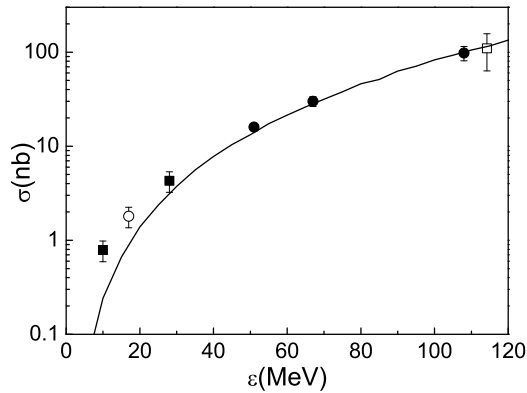}%
\vspace{-0.3cm} \caption{The non-$\phi$ contribution to the $pp \to
pK^+ pK^-$ total cross section versus excess energy $\varepsilon$.
The results of the present calculation are compared with experimental
data from Refs.~\cite{maedaprc77} (closed circles),~\cite{disto}
(open square), \cite{winterplb635} (closed squares),
and~\cite{quenplb515} (open
circle).} \label{tcs}%
\end{figure}

The predictions for the variation of the $pp \to pK^+ p K^-$ total
cross section with excess energy $\varepsilon$, calculated using a
Monte Carlo multiparticle phase-space integration program, are shown
in Fig.~\ref{tcs}. Although the general shape of the experimental
data is described, nevertheless the results very close to threshold
are underestimated. This may be due to the neglect of a $K^+K^-$
final state interaction~\cite{dzyubaplb668}, which might be
associated with the influence of the $a_0$ and $f_0$ scalar
resonances~\cite{maedaprc77}.

\begin{figure}[htdp]
\includegraphics[scale=0.39]{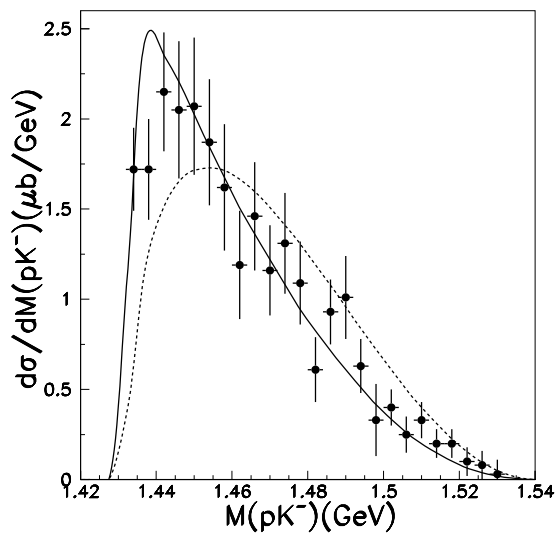}%
\vspace{-0.3cm} \caption{Differential cross section for the $pp \to
pK^+ K^-p$ reaction at the excess energy $\varepsilon = 108$~MeV as
a function of the $K^-p$ invariant mass $M(pK^-)$. The ANKE data of
Ref.~\cite{maedaprc77} are compared to the predictions of the
$N^*(1535)$ model (solid line), whereas the dashed line
represents a normalized four-body phase-space distribution.}
\label{pk108}%
\end{figure}

The predicted $K^-p$ invariant mass spectrum for the $pp \to
pK^+\{K^-p\}$ reaction at $T_p = 2.83$~GeV ($\varepsilon = 108$~MeV)
is compared in Fig.~\ref{pk108} to the experimental data from the
ANKE group~\cite{maedaprc77}. The theoretical model reproduces well
the shape of the data, being much more peaked to lower invariant
masses than the four-body phase-space distribution, which is also
shown. As already indicated in Fig.~\ref{tcs}, the predicted 100~nb
coincides with the experimental value of $(98\pm8\pm15)$~nb, where
the first error is statistical and the second
systematic~\cite{maedaprc77}.

\begin{figure}[htdp]
\includegraphics[scale=0.39]{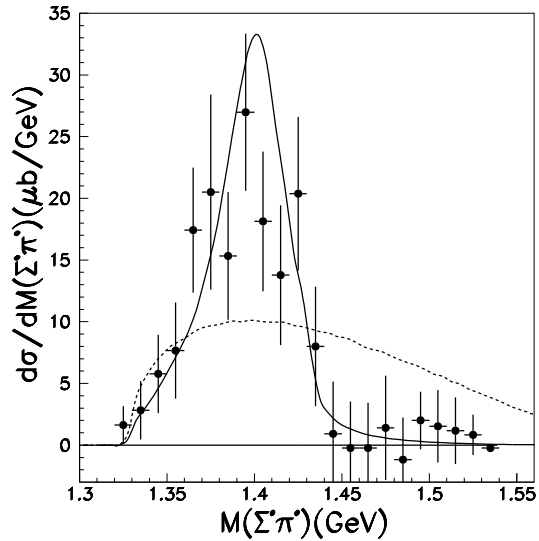}%
\vspace{-0.3cm} \caption{Differential cross section for the $pp \to
pK^+ \pi^0 \Sigma^0$ reaction at an excess energy of $\varepsilon =
212$~MeV. The predictions of the $N^*(1535)$ model (solid
line) have been scaled down by a factor of about 1.5/4 before being
compared to the ANKE data~\cite{zychorplb660}. The fairly
shapeless four-body phase-space distribution (dashed line) has
also been normalized to the total number of experimental events.}
\label{sigmapi212}%
\end{figure}

The corresponding results for the $\pi^0 \Sigma^0$ invariant mass
distribution for the $pp \to pK^+ \pi^0 \Sigma^0$ reaction at the
same beam energy, but excess energy $\varepsilon = 212$~MeV, are
shown in Fig.~\ref{sigmapi212} together with the ANKE
data~\cite{zychorplb660}. Although the statistics are low, the shape
of the spectrum is described correctly, with a rather asymmetric
$\Lambda(1405)$ peak that is strongly influenced by the opening of
the $\bar{K}N$ threshold, that is by the energy dependence of the
$\Lambda(1405)$ width parametrized by Eq.~\eqref{wid1405}. On the
other hand, the overall normalization of the prediction is too high,
giving a cross section of 4.0~$\mu$b compared to an experimental
value of $(1.5\pm 0.3 \pm 0.6)~\mu$b~\cite{zychorplb660}. The
predicted normalization could, of course, be reduced by considering
the initial state interaction but that would then lower also the
value for the $pp \to pK^+ K^-p$ channel.

Many effects cancel out in the estimation of the ratio of the $pp \to
pK^+ K^-p$ to $pp \to pK^+ \pi\Sigma$ total cross sections. These
include initial state distortions and most of the parameters
connected with the $N^*(1535)$. Combining the two experimental
results one finds that, at a proton beam energy of 2.83~GeV,
\begin{equation}
R_{K\pi} = \frac{\sigma(pp \to pK^+ K^-p)}{\sigma(pp \to pK^+\pi^0
\Sigma^0)} = (65 \pm 24) \times 10^{-3}\,,
\end{equation}%
where only non-$\phi$ events have been considered. This is to be
compared with a value of $R_{K\pi} \sim 25 \times 10^{-3}$ obtained
within the framework of the present model. The theoretical
uncertainties are hard to quantify because they reside to a large
extent in the modelling of the low energy $K^-p/\pi^0\Sigma^0$
system~\cite{gal}, which is based upon a limited experimental data
set. In addition there are possibly small contributions from $I=1$
$s$-wave $K^-p$ pairs or, for the higher masses, also some $p$-wave
contributions. In view of the large experimental and theoretical
uncertainties, the good agreement for the $R_{K\pi}$ ratio is very
satisfactory.

%
%

\section{Summary and Conclusions}
\label{conclusions}

The total and differential cross sections for associated strangeness
production in the $pp \to pK^+\{K^-p\}$ and $pp \to
pK^+\{\pi^0\Sigma^0\}$ reactions have been studied in a unified
approach using an effective Lagrangian model. The basic assumptions
are that both the $K^-p$ and $\pi^0\Sigma^0$ systems come from the
decay of the $\Lambda(1405)$ resonance. This state itself results
from the excitation of the $N^*(1535)$ isobar, for which there is
strong evidence for the importance of hidden strangeness components.
Although only pion exchange has been kept in the $pp\to pN^*(1535)$
reaction, our predictions are sensitive to the $N^*(1535)$ production
rate and pion exchange provides a reasonable description of this.
Within the model, the energy dependence of the $pp \to pK^+K^-p$
total cross section is well reproduced, as are the characteristic
$K^-p$ and $\pi^0\Sigma^0$ invariant mass distributions.

Of particular interest is the ratio $R_{K\pi}$ of the $pp \to
pK^+K^-p$ and $pp \to pK^+\pi^0\Sigma^0$ total cross sections because
in the estimation of $R_{K\pi}$ many unknowns drop out. Apart from
initial state distortion, which has been completely neglected in our
work, the details of the $N^*(1535)$ doorway state are largely
irrelevant provided that this state lies well below the
$K^+\Lambda(1405)$ threshold. Thus the very satisfactory prediction
for $R_{K\pi}$ would remain the same if one assumed that the
processes were driven for example by the $N^*(1650)$ isobar. On the
other hand, it is the absolute value of either cross section that
depends upon the $N^*(1535)$ hypothesis and it is the reasonable
description here that gives further weight to the idea of large
$s\bar{s}$ components in this isobar.

The link between $K^-p$ and $\pi^0\Sigma^0$ production could be
established through the use of much low energy data, which led to the
phenomenological separable potential description of the coupled $K^-p
\rightleftarrows \pi^0\Sigma^0$ systems~\cite{gal}. Although this
particular model gives rise to a single $\Lambda(1405)$ pole it is
merely a parametrization of measured scattering data and we cannot
rule out the possibility that similar results would be obtained if
one used a chiral unitary description which requires two
$\Lambda(1405)$ poles~\cite{gengepja34}.

The production of $K\bar{K}$ resonances, such as the $a_0/f_0$
scalars~\cite{pdg2008}, can clearly not contribute to the $pp \to
pK^+\pi^0\Sigma^0$ reaction. Consequently, even if the model
presented here is only qualitatively correct it would suggest that
non-$\phi$ $K^+K^-$ production in $pp \to pK^+K^-p$ is driven
dominantly through the excitation of $K^+$-hyperon pairs rather than
non-strange mesonic resonances.

Further experimental data are needed and some should be available
soon on the $pp \to pK^+\pi^0\Sigma^0$ reaction at the slightly
higher energy of 3.5~GeV from the HADES collaboration~\cite{FAB2010}.
It would, however, be highly desirable to have data on kaon pair
production at a similar energy in order to provide an independent
check on the value of $R_{K\pi}$ and hence on the approach presented
here.

%
%

\begin{acknowledgments}
We wish to thank Xu Cao, A.~Gal, M.~Hartmann, Feng-Kun Guo,
J.~Nieves, N.~Shevchenko, and Bing-Song Zou for useful discussions,
and the CAS Theoretical Physics Center for Science Facilities for
support and hospitality during the initiation of this work. This
research was partially supported by Ministerio de Educaci\'on
``Estancias de movilidad de profesores e investigadores extranjeros
en centros espa\~noles'', Contract No.\ SB2009-0116.
\end{acknowledgments}
%
%


\begin{thebibliography}{99}
%
\bibitem{pdg2008} C.~Amsler \emph{et al.}, Phys.\ Lett.\ B \textbf{667}, 1
(2008).
%
\bibitem{garciaplb582} C.~Garcia-Recio, M.~F.~M.~Lutz and J.~Nieves, Phys.\ Lett.\ B \textbf{582}, 49
(2004).
%
\bibitem{liuprl96} B.~C.~Liu and B.~S.~Zou, Phys.\ Rev.\ Lett.\ \textbf{96}, 042002 (2006);
              B.~C.~Liu and B.~S.~Zou, Phys.\ Rev.\ Lett.\ \textbf{98}, 039102 (2007);
              B.~C.~Liu and B.~S.~Zou, Commun.\ Theor.\ Phys.\ \textbf{46}, 501 (2006).
%
\bibitem{gengprc79} L.~S.~Geng, E.~Oset, B.~S.~Zou and M.~D\"{o}ring, Phys.\ Rev.\ C \textbf{79}, 025203
(2009).
%
\bibitem{BES} J.~Z.~Bai \emph{et al.}, Phys.\ Lett.\ B \textbf{510},
75 (2001); H.~X.~Yang \emph{et al.}, Int.\ J.\ Mod.\ Phys.\ A
\textbf{20}, 1985 (2005).
%
\bibitem{COSY11} P.~Kowina \emph{et al.}, Eur.\ Phys.\ J. A \textbf{22},
293 (2004).
%
\bibitem{Mosel} G.~Penner and U.~Mosel, Phys.\ Rev.\ C \textbf{66},
055211 (2002); \emph{ibid}.\ C \textbf{66}, 055212 (2002);
V.~Shklyar, H.~Lenske and U.~Mosel, Phys.\ Rev.\ C \textbf{72},
015210 (2005).
%
\bibitem{Saghai} B.~Julia-Diaz, B.~Saghai, T.-S.~H.~Lee and F.~Tabakin, Phys.\ Rev.\ C \textbf{73}, 055204
(2006).
%
\bibitem{ELSA} M.~Q.~Tran \textit{et al.}, Phys.\ Lett.\ B \textbf{445},
20 (1998); K.~H.~Glander \textit{et al.}, Eur.\ Phys.\ J.\ A
\textbf{19}, 251 (2004).
%
\bibitem{CLAS} R.~Nasseripour \textit{et al.}, Phys.\ Rev.\ C
\textbf{77}, 065208 (2008).
%
\bibitem{kaisernpa612} N.~Kaiser, T.~Waas and W.~Weise, Nucl.\ Phys.\
A \textbf{612}, 297 (1997).
%
\bibitem{osetprc65} T.~Inoue, E.~Oset and M.~J.~Vicente Vacas, Phys.\ Rev.\ C \textbf{65},
035204 (2002).
%
\bibitem{nievesprd64}J.~Nieves and E.~Ruiz Arriola, Phys.\ Rev.\ D
\textbf{64}, 116008 (2001).
%
\bibitem{doeringepja43} M.~D\"{o}ring and K.~Nakayama, Eur.\ Phys.\
J.\ A \textbf{43}, 83 (2010).
%
\bibitem{dugger} M.~Dugger \textit{et al.}, Phys.\ Rev.\ Lett.\ \textbf{96},
062001 (2006); Phys.\ Rev.\ Lett.\ \textbf{96}, 169905 (2006).
%
\bibitem{caoxuprc78} Xu~Cao and X.~G.~Lee, Phys.\ Rev.\ C \textbf{78},
035207 (2008).
%
\bibitem{doringprc78} M.~D\"{o}ring, E.~Oset and B.~S.~Zou, Phys.\ Rev.\ C \textbf{78}, 025207
(2008).
%
\bibitem{xieprc77} J.~J.~Xie, B.~S.~Zou and H.~C.~Chiang, Phys.\ Rev.\ C \textbf{77}, 015206
(2008).
%
\bibitem{caoxuprc80} Xu~Cao, J.~J.~Xie, B.~S.~Zou and H.~S.~Xu, Phys.\ Rev.\ C \textbf{80}, 025203
(2009).
%
\bibitem{zoureview} B.~S.~Zou, Nucl. Phys. A \textbf{835}, 199 (2010).
%
\bibitem{zhangan} A.~Zhang \textit{et al.}, High Ener. Phys.\ Nucl.\ Phys.\ \textbf{29}, 250
(2005).
%
\bibitem{anepja39} C.~S.~An and B.~S.~Zou, Eur.\ Phys.\ J.\ A
\textbf{39}, 195 (2009).
%
\bibitem{isgurprd18} N.~Isgur and G.~Karl, Phys.\ Rev.\ D \textbf{18},
4187 (1978).
%
\bibitem{dalitz} R.~H.~Dalitz and S.~F.~Tuan, Ann.\ Phys.\ (N.Y.)
\textbf{10}, 307 (1960).
%
\bibitem{inouenpa790} T.~Inoue, Nucl.\ Phys.\ A \textbf{790}, 530
(2007).
%
\bibitem{chiral} E.~Oset and A.~Ramos, Nucl.\ Phys.\ A \textbf{635},
99 (1998); E.~Oset, A.~Ramos and C.~Bennhold, Phys.\ Lett.\ B
\textbf{527}, 99 (2002); J.~A.~Oller and U.-G.~Mei{\ss}ner, Phys.\
Lett.\ B \textbf{500}, 263 (2001); D.~Jido, J.~A.~Oller, E.~Oset and
U.-G.~Mei{\ss}ner, Nucl.\ Phys.\ A \textbf{725}, 181 (2003);
C.~Garcia-Recio, J.~Nieves, E.~Ruiz Arriola and M.~J.~Vicente Vacas,
Phys.\ Rev.\ D \textbf{67}, 076009 (2003); T.~Hyodo, S.~I.~Nam,
D.~Jido and A.~Hosaka, Phys.\ Rev.\ C \textbf{68}, 018201 (2003).
%
\bibitem{zychorplb660} I.~Zychor \textit{et al.}, Phys.\ Lett.\ B
\textbf{660}, 167 (2008).
%
\bibitem{gengepja34} L.~S.~Geng and E.~Oset, Eur.\ Phys.\ J.\ A \textbf{34}, 405
(2007).
%
\bibitem{maedaprc77} Y.~Maeda \textit{et al.}, Phys.\ Rev.\ C \textbf{77},
015204 (2008).
%
\bibitem{WIL09} C.~Wilkin, Acta Phys.\ Polon. Proc.\ Supp.\ \textbf{2},
89 (2009).
%
\bibitem{models}
J.-F.~Germond and C.~Wilkin, Nucl.\ Phys.\ A \textbf{518}, 308
(1990); J.~M.~Laget, F.~Wellers and J.~F.~Lecolley, Phys.\ Lett.\ B
\textbf{257}, 258 (1991); T.~Vetter, A.~Engel, T.~Bir\'o and
U.~Mosel, Phys.\ Lett.\ B \textbf{263}, 153 (1991); E.~Gedalin,
A.~Moalem and L.~Razdolskaja, Nucl.\ Phys.\ A \textbf{650}, 471
(1999); M.~Batini\'c, A.~\v{S}varc and T.-S.~H.~Lee, Physica Scripta
\textbf{56}, 321 (1997); V.~Bernard, N.~Kaiser and U.-G.~Mei{\ss}ner,
Eur.\ Phys.\ J.\ A \textbf{4}, 259 (1999); M.~T.~Pe\~{na},
H.~Garcilazo and D.~O.~Riska, Nucl.\ Phys.\ A \textbf{683}, 322
(2001); G.~F\"aldt and C.Wilkin, Physica Scripta \textbf{64}, 427
(2001); K.~Nakayama, Y.~Oh and H.~Haberzettl, arXiv:0803.3169 (2008).
%
\bibitem {mach} R.~Machleidt, K.~Holinde and C.~Elster, Phys.\ Rep.\ \textbf{149},
1 (1987).
%
\bibitem{tsushima} K.~Tsushima, S.~W.~Huang and A.~Faessler, Phys.\ Lett.\ B \textbf{337}, 245 (1994);
K.~Tsushima, A.~Sibirtsev and A.~W.~Thomas, Phys.\ Lett.\ B
\textbf{39}, 29 (1997).
%
\bibitem{zouprc03} B.~S.~Zou and F.~Hussain, Phys.\ Rev.\ C \textbf{67}, 015204
(2003).
%
\bibitem{feuster} T.~Feuster and U.~Mosel, Phys.\ Rev.\ C
\textbf{58}, 457 (1998); \emph{ibid}. \textbf{59}, 460 (1999).
%
\bibitem{liang} W.~H.~Liang \textit{et al.}, J.\ Phys.\ G \textbf{28}, 333
(2002).
%
\bibitem{gal} N.~V.~Shevchenko, A.~Gal and J.~Mare\v{s}, Phys.\ Rev.\
Lett.\ \textbf{98}, 082301 (2007);  N.~V.~Shevchenko, A.~Gal,
J.~Mare\v{s} and J.~R\'evai, Phys.\ Rev.\ C \textbf{76}, 044004
(2007).
%
\bibitem{pipdata} A.~Baldini, V.~Flamino, W.~G.~Moorhead and D.~R.~O.~Morrison,
        Landolt-B{\"{o}}rnstein, \textit{Numerical Data and Functional Relationships
        in Science an Technology}, vol.~\textbf{12}, ed.\
        H.~Schopper, Springer-Verlag(1988).
%
\bibitem {gill} J.~Gillespie, \textit{Final-State Interactions}, (Holden-Day, San Francisco,
1964).
%
\bibitem{sibiepja06} A.~Sibirtsev, J.~Haidenbauer and U.-G.~Mei{\ss}ner, Eur.\ Phys.\ J.\ A \textbf{27},
263 (2006).
%
\bibitem{dzyubaplb668} A.~Dzyuba \textit{et al.}, Phys.\ Lett.\ B
\textbf{668}, 315 (2008).
%
\bibitem{disto} F.~Balestra \textit{et al.}, Phys.\ Rev.\ C \textbf{63}, 024004
(2001).
%
\bibitem{winterplb635} P.~Winter \textit{et al.}, Phys.\ Lett.\ B
\textbf{635}, 23 (2006).
%
\bibitem{quenplb515} C.~Quentmeier \textit{et al.}, Phys.\ Lett.\ B \textbf{515},
276 (2001).
%
\bibitem{FAB2010} L.~Fabbietti and E.~Epple, Nucl.\ Phys.\ A
\textbf{835}, 333 (2010).
%
\end{thebibliography}
\end{document}